\documentclass[pre,twocolumn,floatfix,amsmath,amsfonts,longbibliography]{revtex4}
\usepackage{graphicx,epstopdf,url}
\usepackage{color}
\usepackage[colorlinks=true,urlcolor=blue,citecolor=blue,linkcolor=blue,urlcolor=blue]{hyperref}
\usepackage[active]{srcltx}

\begin{document}
\title{
Wrinkles in graphene suspended on flat substrates: structure and collapse under hydrostatic pressure
}
\author{Alexander V. Savin}
\email[]{asavin@chph.ras.ru}
\affiliation{
N.N. Semenov Federal Research Center for Chemical Physics of the Russian Academy of Sciences,
4 Kosygin St., Moscow 119991, Russia}
\affiliation{
Plekhanov Russian University of Economics, 36 Stremyanny Lane, Moscow 117997, Russia
}
\author{Artem P. Klinov}
\email[]{aklinov@chph.ras.ru}
\affiliation{
N.N. Semenov Federal Research Center for Chemical Physics of the Russian Academy of Sciences,
4 Kosygin St., Moscow 119991, Russia}

\date{\today}

\begin{abstract}
The method of molecular dynamics and molecular mechanics has been used to numerically simulate
the formation of wrinkle systems during compression of a graphene sheet lying on a flat solid substrate.
It is shown that under uniaxial compression the nanosheet can transition into several stable wrinkled states: the most energetically favorable one is a linear wrinkle of infinite length. Higher energy states include wrinkles of finite length aligned along the same line where their ends partially overlap.
Under biaxial compression, the graphene nanosheet can contain one linear wrinkle or two linear non-intersecting or intersecting wrinkles corresponding to weak, medium and strong compression, respectively.
There are several nanosheet states with intersecting wrinkles that differ in structure of intersection area.
The effect of external hydrostatic pressure on the shape of wrinkles has been studied.
It is shown that there is a critical pressure value at which the wrinkle either completely
flattens (disappears) or collapses into a vertical two-layer fold (the first scenario is possible only with weak compression of the sheet).
\end{abstract}


\maketitle

\section{Introduction \label{sec1}}

Carbon atoms are capable of creating numerous structures such as
monoatomic crystalline layer of graphene that has recently attracted much attention from researchers
\cite{Novoselov2004,Geim2007,Soldano2010,Baimova2014,Baimova2014a}.
This nanomaterial is of interest because of its unique electronic \cite{Geim2009}, mechanical \cite{Lee2008} and thermal \cite{Baladin2008,Liu2015} properties.

A popular method for producing graphene is the chemical vapour deposition (CVD) method,
in which graphene is grown on a substrate in a carbon-rich environment.
The CVD method often leads to the appearance of topological defects (during the cooling process, the graphene sheet undergoes out-of-plane bending),
such as ripples \cite{Tapaszto2012} and wrinkles \cite{Zhu2012}.
Defects of this type can be formed due to the roughness of the substrate \cite{Lui2009} and difference between  thermal expansion coefficients of graphene and the substrate \cite{Obraztsov2007}.
The presence of such defects can change the properties of graphene: electrical conductivity \cite{Zhu2012, Guan2024},
thermal conductivity \cite{Chen2012, Wang2014} and elasticity \cite{Wang2011}.
The structures of wrinkles and folds appearing on the sheet can be used as channels
for the injection and storage of liquid between graphene and its substrate \cite{Carbone2019},
as well as for its spatially selective chemical functionalization \cite{Deng2019}.
Therefore, understanding the laws of wrinkle and fold formation and explaining the mechanisms of their interactions
is important for creating graphene-based nanodevices.

Out-of-plane (transverse) deformations of graphene can be divided into ripples (corrugations), wrinkles and folds
depending on their physical dimensions and topology \cite{Deng2016,Deng2018}.
Such deformations in graphene can occur for various reasons.
For example, when using the chemical vapour deposition (CVD) method, wrinkles and folds are formed
as a result of thermal compression of the substrate at the cooling stage \cite{Deng2016,Wang2021,Pan2011,Zhao2024}.
Out-of-plane defects may appear during the graphene sheet transfer procedure \cite{Lanza2013,Liu2011}.

Individual wrinkles and folds have been previously described by quasi-analytical models based on calculus of variations
 \cite{Zhu2012,Zhang2013a,Cox2015,Aljedani2020,Cox2020,Aljedani2021,Aljedani2021a},
continuum mechanics models using the finite element method
\cite{Zhang2013,Zhang2014} and all-atom models using molecular dynamics
\cite{Guo2013,Mulla2015,Li2016,Savin2019,Zhu2020,Zhao2020}.
Note that quasi-analytical models allow one to describe only wrinkles and folds that occur
during uniaxial compression. With continuum mechanics treatment it is possible to consider
uni- and biaxial compression of sufficiently large graphene sheets (larger than $500\times$500~nm$^2$).
However, in these works nanosheet compressions are less than 2.4\%, when only ripples and wrinkles are formed as a result of compression, and the influence of thermal fluctuations is not taken into account.
To overcome these limitations, the molecular dynamics method can be applied despite it is currently possible to model only sheets of size $50\times50$~nm$^2$ at times of several nanoseconds.

In this work, the methods of molecular dynamics and molecular mechanics will be used to numerically simulate the formation of wrinkle systems in graphene sheet lying on a flat rigid substrate under uniaxial and biaxial compression.
It will be shown that at uniaxial compression of a sheet, in addition to a linear wrinkle of infinite length, higher-energy stable bound states of wrinkles of finite length can also be formed.
These wrinkles lie along one line and their ends partially overlap.
With biaxial compression, the sheet can have stationary states with one linear wrinkle, with two linear orthogonal non-intersecting and intersecting wrinkles (these types of wrinkles are energetically favourable under weak, medium and strong compression, respectively).
States with orthogonal intersecting wrinkles can have different shapes in the region of their intersection.

Special attention will be paid to the effect of external hydrostatic pressure on the shape of wrinkles.
It will be shown that there is always a critical pressure value at which the wrinkle either completely flattens (disappears), or collapses into a denser form of a vertical two-layer fold (the first scenario is possible only with weak compression of the sheet).
It should be noted that previously only the collapse of hollow carbon nanotubes under pressure was studied \cite{Elliott2004,Gadagkar2006,Umeno2019,Magnin2021,Hu2024}.

In Sec.~\ref{sec2}, the force field and modeling methods are described.
The stability of a flat uniformly compressed sheet state is analyzed in Sec.~\ref{sec3}.
The structures of wrinkles formed during uniaxial and biaxial compression of the sheet are described in Sec.~\ref{sec4}
and \ref{sec5}. The stability of the systems of stationary wrinkles of a compressed sheet to thermal fluctuations
is tested in Sec.~\ref{sec6}. In Sec.~\ref{sec7}, the effect of pressure on the shape of wrinkles is described
and their collapse under pressure is modeled. Section \ref{sec8} concludes our work.
\begin{figure}[tb]
\begin{center}
\includegraphics[angle=0, width=1.0\linewidth]{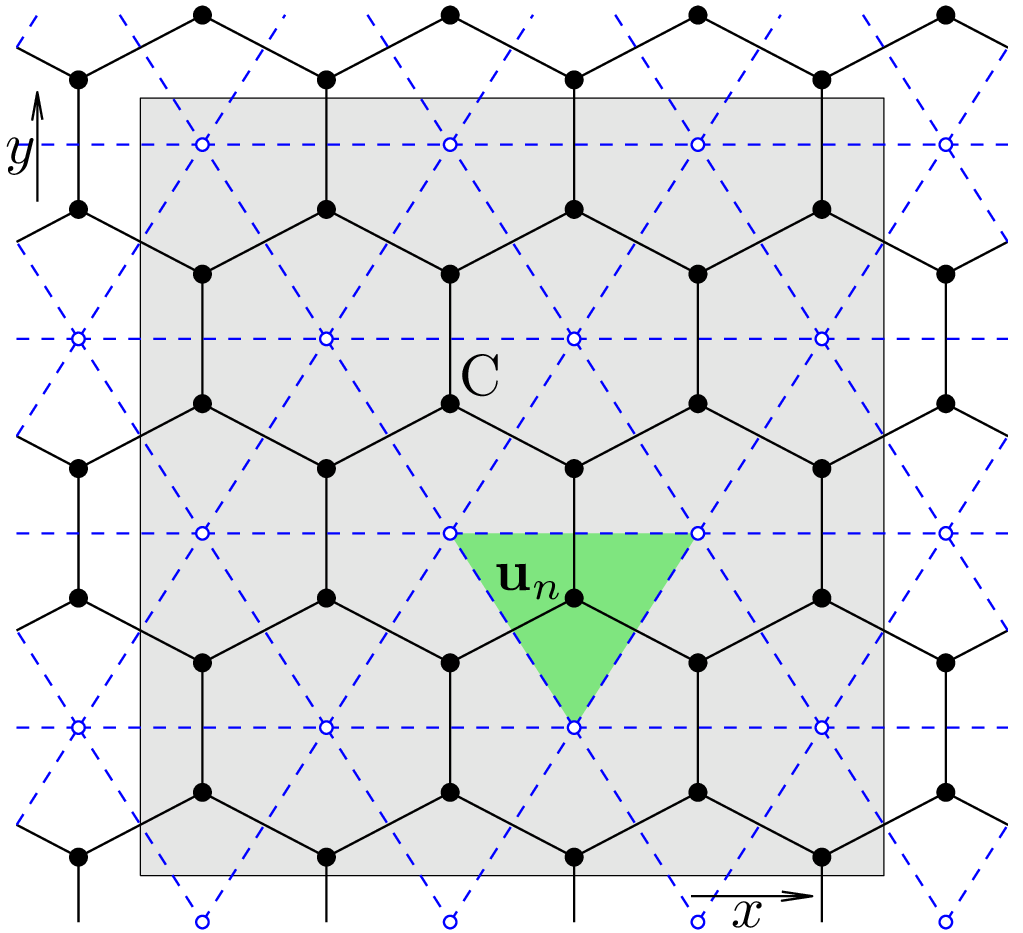}
\end{center}
\caption{\label{fg01}\protect
A rectangular graphene sheet with $N_x=6$, $N_y=8$. Black dotes denote carbon atoms C, black solid lines -- valent bonds C--C. Gray zone depicts a unit cell.
Small light disks show the positions of the geometric centers of the hexagons of valence bonds,
dotted lines show the triangulation of a graphene sheet using these centers -- the surface
of the sheet is divided into triangles with one carbon atom in each (the corresponding triangle to the $n$-th atom is highlighted, ${\bf u}_n$ is a vector of coordinates of this atom).
}
\end{figure}

\section{Model \label{sec2}}

Consider a rectangular graphene sheet (see Fig.~\ref{fg01})
whose chains of valence bonds along the $x$ axis have
a "zig-zag" structure with $N_x$ atoms, and along the $y$ axis -- an "armchair" structure with $N_y$ atoms (the sheet lies parallel
to the plane $z=0$). We introduce periodic boundary conditions at the edges of the sheet. To satisfy them, we take $N_x$
as a multiple of two, and $N_y$ -- as a multiple of four. In this case nanosheet consists of  $N=N_xN_y/2$ carbon atoms and has size $A_x^0\times A_y^0$,
where period along $x$-axis is $A_x^0=N_xr_{c}\sqrt{3}/2$, and period along $y$-axis $A_y^0=3N_yr_{c}/4$
($r_{c}=1.418$~\AA~ -- length of C--C bond).
In all calculations we take graphene nanosheet of size $N_x=416$, $N_y=480$, the number of carbon atoms $N=208\times 480=99840$, period along $x$-axis $A_x^0=51.086$,
period along $y$-axis $A_y^0=51.048$~nm.

The Hamiltonian of the nanosheet:
\begin{equation}
H=\sum_{n=1}^N\left[\frac12M(\dot{\bf u}_n,\dot{\bf u}_n)+Q_n+W_n+Z({\bf u}_n)\right],
\label{f1}
\end{equation}
where $M$ is the mass of a carbon atom, ${\bf u}_n=(x_n(t),y_n(t),z_n(t))$ is a vector defining
the position of $n$-th atom at time $t$.
The first term of the sum (\ref{f1}) is the kinetic energy, mass $M=12m_p$ ($m_p$ -- the proton mass).
\begin{table}[tb]
\caption{
Critical values of dimensionless uniaxial $d_1$ and biaxial $d_2$ compression parameter at different values of temperature $T$ and pressure $P$ (adhesion energy $\varepsilon_0=0.075$~eV).
\label{tab1}
}
\begin{center}
\begin{tabular}{c|cc|cc|cc}
\hline\hline
 $P$  & ~$T=150$ & K~~~  & ~$T=300$ & K~~~  & ~$T=600$ & K~~~   \\
 (GPa)   & $d_1$ & ~~~$d_2$~~~ & $d_1$ & ~~~$d_2$~~~ & $d_1$ & $d_2$ (\%) \\
 \hline
   0  & 4.2 & 2.4 & 4.1 & 2.4 & 4.1 & 2.4 \\
0.16  & 4.4 & 2.5 & 4.3 & 2.5 & 4.1 & 2.6 \\
0.32  & 4.6 & 2.6 & 4.5 & 2.6 & 4.4 & 2.7 \\
0.48  & 4.8 & 2.7 & 4.7 & 2.7 & 4.6 & 2.8 \\
0.64  & 4.9 & 2.8 & 4.8 & 2.8 & 4.7 & 2.9 \\
0.80  & 5.1 & 2.9 & 4.9 & 2.9 & 5.0 & 3.0 \\
\hline\hline
\end{tabular}
\end{center}
\end{table}
\begin{table}[tb]
\caption{
Critical values of dimensionless uniaxial $d_1$ and biaxial $d_2$ compression parameter at different values of adhesion energy $\varepsilon_0$ and external pressure $P$ (temperature $T=600$~K).
\label{tab2}
}
\begin{center}
\begin{tabular}{c|cc|cc|cc}
\hline\hline
 $P$  & $\varepsilon_0=0.032$ &eV~~~  & $\varepsilon_0=0.075$ &eV~~~  & $\varepsilon_0=0.15$ &eV~~\\
 (GPa)   & $d_1$ & $d_2$ & $d_1$ & $d_2$ & $d_1$ & $d_2$ (\%)\\
 \hline
   0 & 3.0 & 1.7 & 4.1 & 2.4 & 5.7 & 3.3 \\
0.16 & 3.2 & 2.0 & 4.1 & 2.6 & 5.8 & 3.4 \\
0.32 & 3.5 & 2.1 & 4.4 & 2.7 & 6.0 & 3.5 \\
0.48 & 3.7 & 2.2 & 4.6 & 2.8 & 6.1 & 3.6 \\
0.64 & 4.0 & 2.4 & 4.7 & 2.9 & 6.2 & 3.6 \\
0.80 & 4.2 & 2.5 & 5.0 & 3.0 & 6.4 & 3.7 \\
\hline\hline
\end{tabular}
\end{center}
\end{table}
\begin{figure*}[tb]
\begin{center}
\includegraphics[angle=0, width=1.0\linewidth]{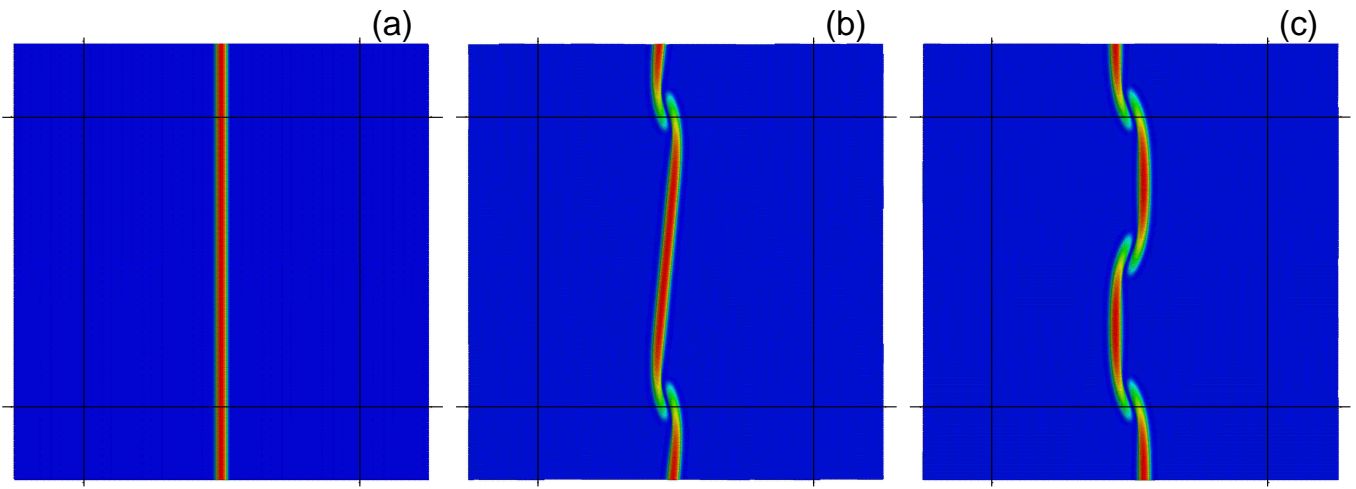}
\end{center}
\caption{\label{fg02}\protect
Stationary states of a uniaxially compressed graphene sheet with (a) one vertical continuous wrinkle of infinite length, (b) with bonded wrinkles of finite length $L=A_y^0$ and (c) $L=A_y^0/2$ (compression along the $x$-axis $d=0.05$).
The top view is presented, the color denotes the amplitude of the vertical displacements of the atoms of the sheet, the blue color corresponds to the zero displacement, the red color corresponds to the maximum (the amplitude of the displacements $A_z=2.028$, 2.053 and 2.023~nm, respectively).
The black lines show the boundaries of periodic cells.
The energy of interaction with the substrate $\epsilon_0=0.075$~eV, pressure $P=0$.
}
\end{figure*}

The second term of the sum (\ref{f1}) $Q_n$ describes the valent interaction of the $n$-th atom with the other atoms of the graphene sheet. It includes deformation energy of valent bonds, valent and torsional angles -- a detailed description of force field is presented in Ref. \cite{Savin2010}.
The valence bond between two neighboring carbon atoms $n$  and  $k$  can  be  described
by the Morse potential
\begin{equation}
U_1({\bf u}_n,{\bf u}_k)=\epsilon_{c}\left[e^{-\alpha(r_{nk}-r_{c})}-1\right]^2,
\label{f2}
\end{equation}
where $r_{nk}=|{\bf u}_n-{\bf u}_k|$ is distance between atoms ,  $\epsilon_{c}=4.9632$~eV is bond energy.
\begin{table}[tb]
\caption{
Characteristic values of uniaxial compression of a graphene sheet $\alpha_1$, $\alpha_2$, $\alpha_3$, $d_1$
at different values of adhesion energy $\varepsilon_0$
(pressure $P=0$).
\label{tab3}
}
\begin{center}
\begin{tabular}{c|cccc}
\hline\hline
~~$\varepsilon_0$ (eV)~~  & $\alpha_1$ & $\alpha_2$  & $\alpha_3$ & $d_1$ \\
 \hline
0.032 & ~~0.0140~ & ~0.0150~ & ~0.0155~ & ~0.030~ \\
0.075 & ~0.0185 & 0.0205 & 0.0214 & 0.041 \\
0.150 & ~0.0240 & 0.0260 & 0.0281 & 0.057 \\
\hline\hline
\end{tabular}
\end{center}
\end{table}

Valence angle deformation energy between three adjacent carbon atoms $n$, $k$, and $l$ can be described
by the potential
\begin{equation}
U_2({\bf u}_n,{\bf u}_k,{\bf u}_l)=\epsilon_\varphi(\cos\varphi-\cos\varphi_0)^2,
\label{f3}
\end{equation}
where $\cos\varphi=({\bf u}_n-{\bf u}_k,{\bf u}_l-{\bf u}_k)/r_{nk}r_{kl}$,
and $\varphi_0=2\pi/3$ is the equilibrium valent angle. Parameters $\alpha=17.889$~nm$^{-1}$
and $\epsilon_\varphi=1.3143$~eV can be found from the small amplitude oscillations spectrum
of the graphene sheet \cite{Savin2008}.

Valence bonds between four adjacent carbon atoms $n$, $m$, $k$, and $l$ constitute torsion angles,
the potential energy of which can be defined as
\begin{equation}
U_3({\bf u}_n,{\bf u}_m,{\bf u}_k,{\bf u}_l)=\epsilon_\phi(1-\cos\phi),
\label{f4}
\end{equation}
where $\phi$ is the corresponding torsion angle ($\phi=0$ is the equilibrium value of the angle)
and $\epsilon_\phi=0.499$~eV.

The third term $W_n$ of the sum (\ref{f1}) corresponds to non-valent interactions of carbon atoms 
described by the Lennard-Jones potential (6,12) \cite{Setton1996}
\begin{equation}
V(r)=4\epsilon_0[(\sigma/r)^{12}-(\sigma/r)^6],
\label{f5}
\end{equation}
where $r$ is distance between atoms,
$\epsilon_0=0.002757$~eV is the binding energy and $\sigma=3.393$~\AA.
The potential attains a minimum value of $-\epsilon_0$ at $r_0=2^{1/6}\sigma=3.807$~\AA~
(equilibrium interaction distance).

The last term describes the interaction of atoms of a sheet with a flat substrate on which it lies. Let the flat substrate fill the half-space $z\le 0$. In this case, the interaction energy of the carbon atom with the substrate can be described by Lennard-Jones potential (3,9)
\cite{Zhang2013,Zhang2014,Aitken2010}
\begin{equation}
Z({\bf u})=Z(z)=\varepsilon_0[(h_0/z)^9-3(h_0/z)^3]/2,
\label{f6}
\end{equation}
where $\varepsilon_0$ is interaction energy (adhesion energy) and $h_0$ is equilibrium distance to the surface of half-space.
For silicon oxide substrate SiO$_2$ energy  $\varepsilon_0=0.074$~eV,
distance $h_0=5$~\AA~ \cite{Koenig2011}. For the surface of the ice crystal adhesion energy $\varepsilon_0=0.029$~eV
\cite{Savin2019a}, and for the surface (111) of the nickel crystal $\varepsilon_0=0.125$~eV \cite{Giovannetti2008,Kozlov2012}.
In the following, we will use the three most characteristic values $\varepsilon_0=0.032$, 0.075 and 0.15~eV
to simulate the weak, medium and strong interaction of the sheet with the substrate, respectively.
\begin{figure}[tb]
\begin{center}
\includegraphics[angle=0, width=1.0\linewidth]{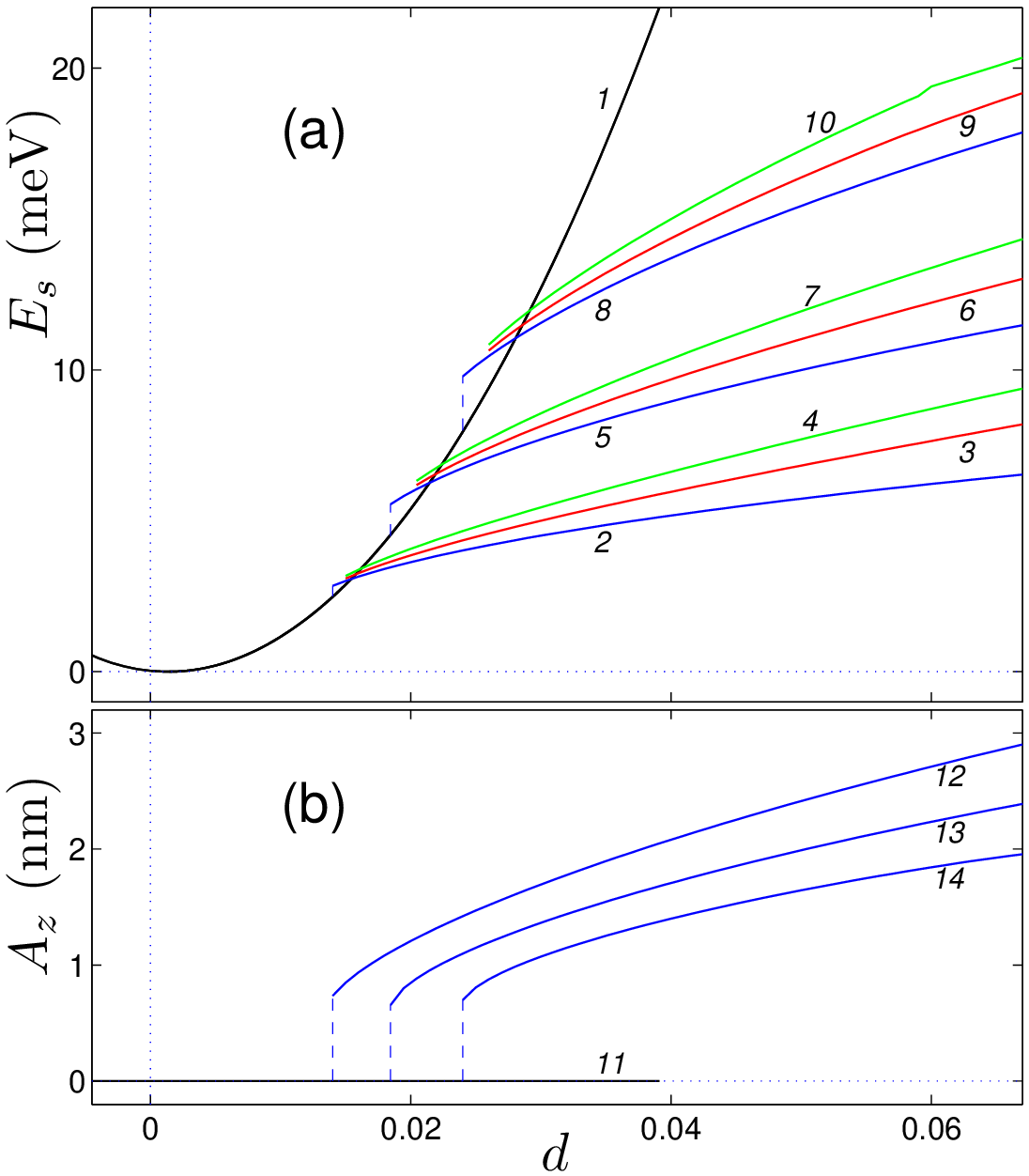}
\end{center}
\caption{\label{fg03}\protect
Dependence of (a) the specific energy of the stationary state of a uniaxial compressed graphene sheet $E_s=(E-E_0)/N$ and (b) the maximum amplitude of vertical displacements of sheet atoms $A_z$ on dimensionless coefficient
of uniaxial compression $d$: for a flat homogeneous compressed state (curves 1, 11);
for one continuous wrinkle (curves 2, 5, 8, 12, 13, 14); for wrinkles of finite length $L=A_y^0$ (curves 3, 6, 9) and for wrinkles of finite length $L=A_y^0/2$ (curves 4, 7, 10).
Curves 2, 3, 4, 12 were obtained at adhesion energy $\epsilon_0=0.032$~eV;
curves 5, 6, 7, 13 -- at $\epsilon_0=0.075$~eV; curves 8, 9, 10, 14 -- at $\epsilon_0=0.15$~eV.
}
\end{figure}

The dynamics of a thermalized graphene sheet under external hydrostatic pressure $P$
is described by a system of Langevin equations
\begin{equation}
M\ddot{\bf u}_n=-\frac{\partial H}{\partial{\bf u}_n}-\Gamma M\dot{\bf u}_n+\Xi_n+PS_n{\bf e}_n,~n=1,...,N,
\label{f7}
\end{equation}
where $\Gamma=1/t_r$ is friction coefficient characterizing the intensity of energy exchange with thermostat
(relaxation time $t_r=10$~ps), $\Xi_n=\{\xi_{n,i}\}_{i=1}^3$ is the three-dimensional vector of normally
distributed random forces normalized by the following conditions:
$$
\langle\xi_{n,i}(t)\xi_{k,j}(s)\rangle=2\Gamma Mk_BT\delta_{nk}\delta_{ij}\delta(s-t),
$$
where $T$ is the thermostat temperature, $k_B$ is the Boltzmann constant.
The vector $P S_n{\bf e}_n$ in Eq. (\ref{f7}) denotes force exerted by hydrostatic pressure $P$
acting orthogonally to the surface of the sheet.
Here $S_n$ denotes the area of the triangle formed by the centers of neighboring hexagons (see Fig.~\ref{fg01}), ${\bf e}_n$ -- the unit vector of the normal to the triangle plane.
The geometric centers of the valence bond hexagons allow us to triangulate the surface of the sheet by them, so that in the center of each triangle there is one carbon atom, and the sum of the areas of the triangles equals the surface area.
\begin{table}[tb]
\caption{
Characteristic values of biaxial compression of a graphene sheet $\beta_1$, $\beta_2$, $\beta_3$, $d_2$
at different values of adhesion energy $\varepsilon_0$
(pressure $P=0$).
\label{tab4}
}
\begin{center}
\begin{tabular}{c|cccc}
\hline\hline
 ~~$\varepsilon_0$ (eV)~~  & $\beta_1$ & $\beta_2$  & $\beta_3$ & $d_2$ \\
 \hline
0.032 & ~~0.0101~ & ~0.0140~ & ~0.0275~ & ~0.017~ \\
0.075 & ~0.0141 & 0.0195 & 0.0375 & 0.024 \\
0.150 & ~0.0184 & 0.0252 & 0.0510 & 0.033 \\
\hline\hline
\end{tabular}
\end{center}
\end{table}

\section{Loss of stability of the flat state of the sheet during its uniform compression \label{sec3}}

Let us simulate the dynamics of a graphene sheet lying on a substrate $z=0$,
under uniform compression in the plane $xy$.
To do this, we numerically integrate the system of equations of motion (\ref{f7}) with the initial conditions
\begin{eqnarray}
\nonumber
x_n(0)=(1-d_x) x_n^0,~y_n(0)=(1-d_y) y_n^0,~z_n(0)=h_0,\\
\dot{x}_n(0)=0,~~\dot{y}_n(0)=0,~~\dot{z}_n(0)=0,~~ \label{f8} \\
n=1,2,...,N,~~ \nonumber
\end{eqnarray}
and with unit cell dimensions $A_x=(1-d_x)A_x^0$, $A_y=(1-d_y)A_y^0$ [square of the compressed unit cell $A_x\times A_y=(1-d_x)(1-d_y)A_x^0,A_y^0$].
Here vectors $\{{\bf u}_n^0= (x_n^0,y_n^0,h_0)\}_{n=1}^N$ denote the ground state of an uncompressed nanosheet with cell dimensions $A_x^0$, $A_y^0$. The coefficient $0\le d_x<1$ determines nanosheet compression along $x$-axis and the coefficient $0\le d_y<1$ -- along $y$-axis. Under uniaxial compression
$d_x=d>0$, $d_y=0$, under biaxial compression $d_x=d_y=d>0$.
\begin{figure*}[tb]
\begin{center}
\includegraphics[angle=0, width=1.0\linewidth]{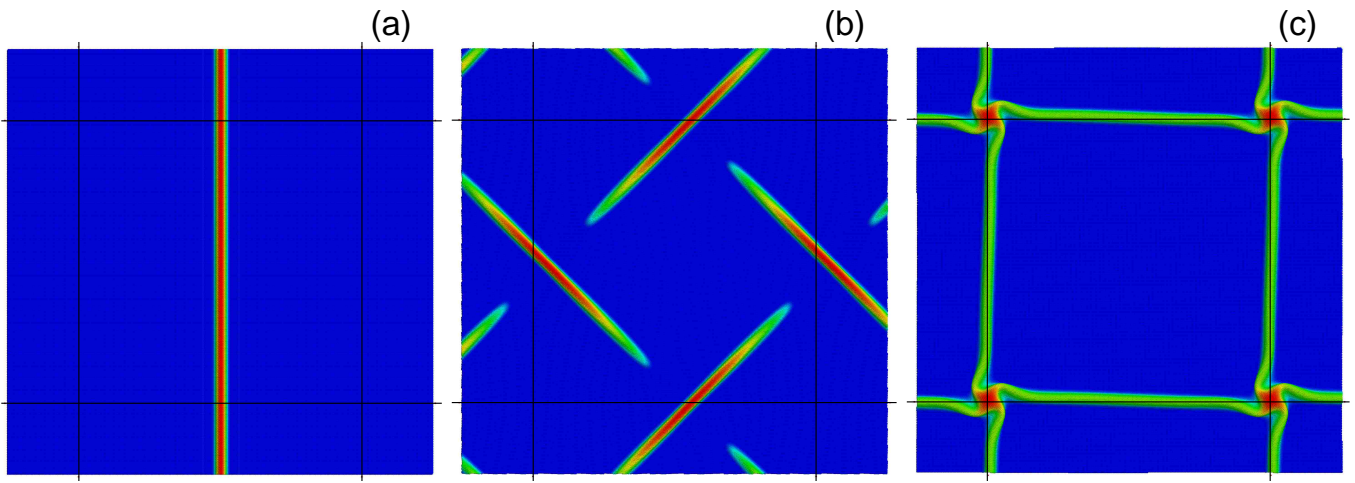}
\end{center}
\caption{\label{fg04}\protect
Stationary states of a biaxially compressed graphene sheet (a) at compression $d=0.018$
(one linear wrinkle, amplitude of out-of-plane displacements $A_z=1.333$~nm),
(b) at $d=0.030$ (two orthogonal non-intersecting wrinkles, $A_z=1.864$~nm),
(c) at $d=0.050$ (two orthogonal intersecting wrinkles, $A_z=3.307$~nm).
Adhesion energy $\varepsilon_0=0.075$~eV, pressure $P=0$.
}
\end{figure*}

The equations of motion Eq. (\ref{f7}) are solved numerically using the velocity Verlet method
\cite{Verlet1967}. A time step of 1 fs is used in the simulations, since further reduction of the time
step has no appreciable effect on the results.

Numerical modeling has shown that for each value of temperature and pressure there are critical values of $d_1>0$ for uniaxial and $d_2>0$ for biaxial compression.
Under compression $d<d_i$ ($i=1,2$), the sheet retains its initial flat shape (only low-amplitude thermal vibrations of atoms are present), while at $d>d_i$ wrinkles and folds are formed in the direction away from the substrate.

The dependence of the critical compression values $d_1$, $d_2$ on temperature $T$ and pressure $P$
is shown in Table. \ref{tab1}.
As can be seen from the table, at zero pressure, the critical compression values are practically
independent of the temperature value. Only at high pressure $P>0.16$~GPa, an increase in temperature
leads to a slight decrease in the values of $d_1$, $d_2$. An increase in pressure stabilizes the flat shape of the sheet ($d_i\nearrow$ at $P\nearrow$).

The dependence of the critical compression values on the interaction energy of the sheet with the substrate $\varepsilon_0$
and the pressure $P$ is shown in Table. \ref{tab2}. As can be seen from the table, an increase in the interaction energy
leads to a significant increase in the critical compression values.
This is especially noticeable at pressure of $P=0$. In the absence of pressure, doubling $\varepsilon_0$ leads
to a 1.4-fold increase in the critical compression values of $d_1$ and $d_2$.

\section{Stationary states of a uniaxially compressed sheet \label{sec4}}

In the absence of pressure (at $P=0$), the stationary state of a compressed graphene sheet can be found as a solution of minimization problem
\begin{equation}
E=\sum_{n=1}^N [Q_n+W_n+Z({\bf u}_n)]\rightarrow\min,
\label{f9}
\end{equation}
with fixed unit cell dimensions $A_x=(1-d_x)A_x^0$, $A_y=(1-d_y)A_y^0$.

Problem of finding minimum (\ref{f9}) was solved numerically by conjugate gradients method \cite{Fletcher1964,Shanno1976}.
The nanosheet configurations obtained by integrating the system of equations of motion were used as starting points (\ref{f7}).
The resulting stationary state of the compressed sheet $\{ {\bf v}_n=(x_n,y_n,z_n)\}_{n=1}^N$ is convenient to characterize by its specific energy $E_s=(E-E_0)/N$, where $E_0$ -- the energy of the ground (uncompressed) state $\{ {\bf u}_n^0\}_{n=1}^N$, and by the amplitude of the transverse displacements of the atoms of the sheet $A_z=\max_n(z_n-h_0)$.
\begin{figure*}[tb]
\begin{center}
\includegraphics[angle=0, width=1.0\linewidth]{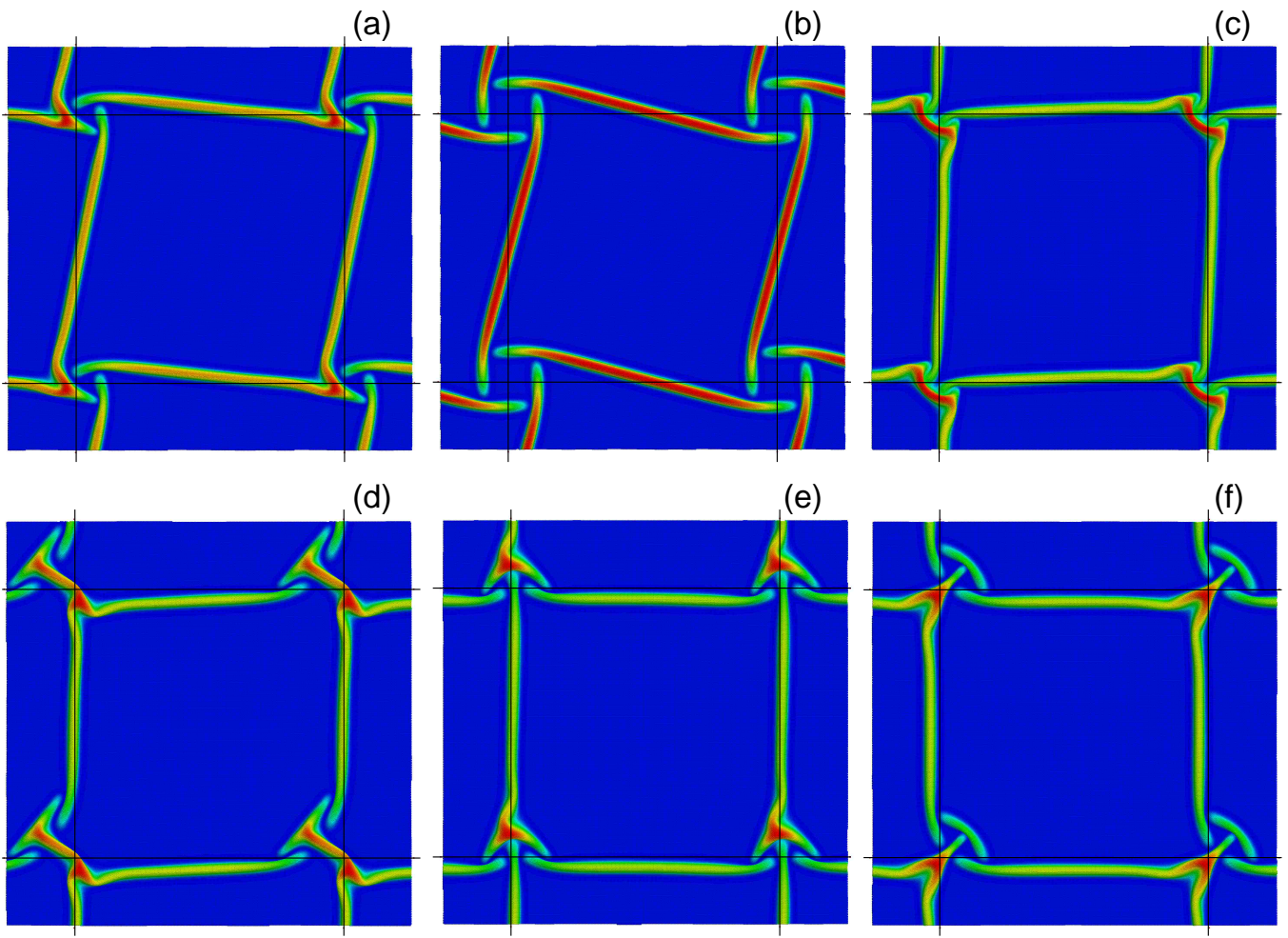}
\end{center}
\caption{\label{fg05}\protect
Stationary states of a biaxially compressed graphene sheet with two orthogonal
wrinkles having different topologies in the intersection area at compression $d=0.05$:
(a) structure with specific energy $E_s=21.03$~meV and amplitude of out-of-plane dispacements $A_z=2.54$~nm; (b) 21.08~meV, 2.13~nm; (c) 21.33~meV, 2.86~nm;
(d) 21.57~meV, 2.92~nm; (e) 21.64~meV, 3.05~nm; (f) 21.68~meV, $2.97$~nm.
Adhesion energy $\varepsilon_0=0.075$~eV, pressure $P=0$.
}
\end{figure*}

There are several solutions of the problem (\ref{f9}) for uniaxial compression along the $x$ axis ($d_x=d$, $d_y=0$). In addition to a stable flat homogeneous compressed state at $d<d_1$, there are three more types of solutions:
a state with one continuous vertical wrinkle (effective wrinkle length $L=\infty$)
and states with adjacent wrinkles of finite length $L=A_y^0$ and $A_y^0/2$
-- see Fig. \ref{fg02}. Here, in addition to one vertical wrinkle crossing the entire sheet,
a system of several interacting vertical wrinkles of finite length can arise.
The interaction of wrinkles occurs due to their partially overlapping ends, but
a flat area in the form of a narrow valley is always preserved between the ends
-- see Fig. \ref{fg02} (b) and (c).

There are three characteristic values of uniaxial compression $0<\alpha_1<\alpha_2<\alpha_3<d_1$.
A stationary state of a sheet with a single wrinkle of length $L=\infty$ can only exist
at $d>\alpha_1$,
states with interacting wrinkles of finite length -- at $d>\alpha_2$.
The state with a wrinkle of infinite length becomes the most energetically favorable at $d>\alpha_3$.
The dependence of these characteristic values of uniaxial compression on adhesion energy $\varepsilon_0$ is presented in the Table \ref{tab3}. As can be seen from the table,
an increase in the energy of interaction with the substrate (an increase in the adhesion energy) leads to a monotonous
increase in these compression values.

The dependence of the specific energy $E_s=(E-E_0)/N$ and the amplitude of the transverse displacements $A_z$ for these
four types of stationary states of a graphene sheet on dimensionless compression $d$ is
shown in Fig. \ref{fg03}.
A stationary flat uniformly compressed state exists only when $d<d_1$. In this case, the energy
of the state increases in proportion to the square of compression ($E_s\propto d^2$ when $d\nearrow$),
and the amplitude of the transverse displacements is always zero ($A_z\equiv 0$).
A state with one continuous wrinkle exists only at $d>\alpha_1$,
and a configuration of adjacent wrinkles of finite length exists at $d>\alpha_2$.
The energy of the wrinkled state grows as the square root of compression ($E_s\propto d^{1/2}$ at $d\nearrow$).
At $d<\alpha_3$ a flat state is more energetically favorable, while at $d>\alpha_3$ -- it is the state with one continuous wrinkle.
The amplitude of the transverse displacements $A_z$ for wrinkled states increases monotonously with increasing compression. 
An increase in the energy of interaction with the substrate leads to increase in the energy of wrinkled state and to reduction of the wrinkle height.
The energy of a state with several wrinkles is always higher than the energy of a state with one wrinkle.

Note that the configuration with one continuous infinite wrinkle has been considered by many
authors -- see for example \cite{Aljedani2020,Cox2020,Zhang2013,Guo2013,Savin2019}.
To the best of our knowledge, the nanosheet with bonded wrinkles of finite length has not yet been studied.
\begin{figure}[tb]
\begin{center}
\includegraphics[angle=0, width=1.0\linewidth]{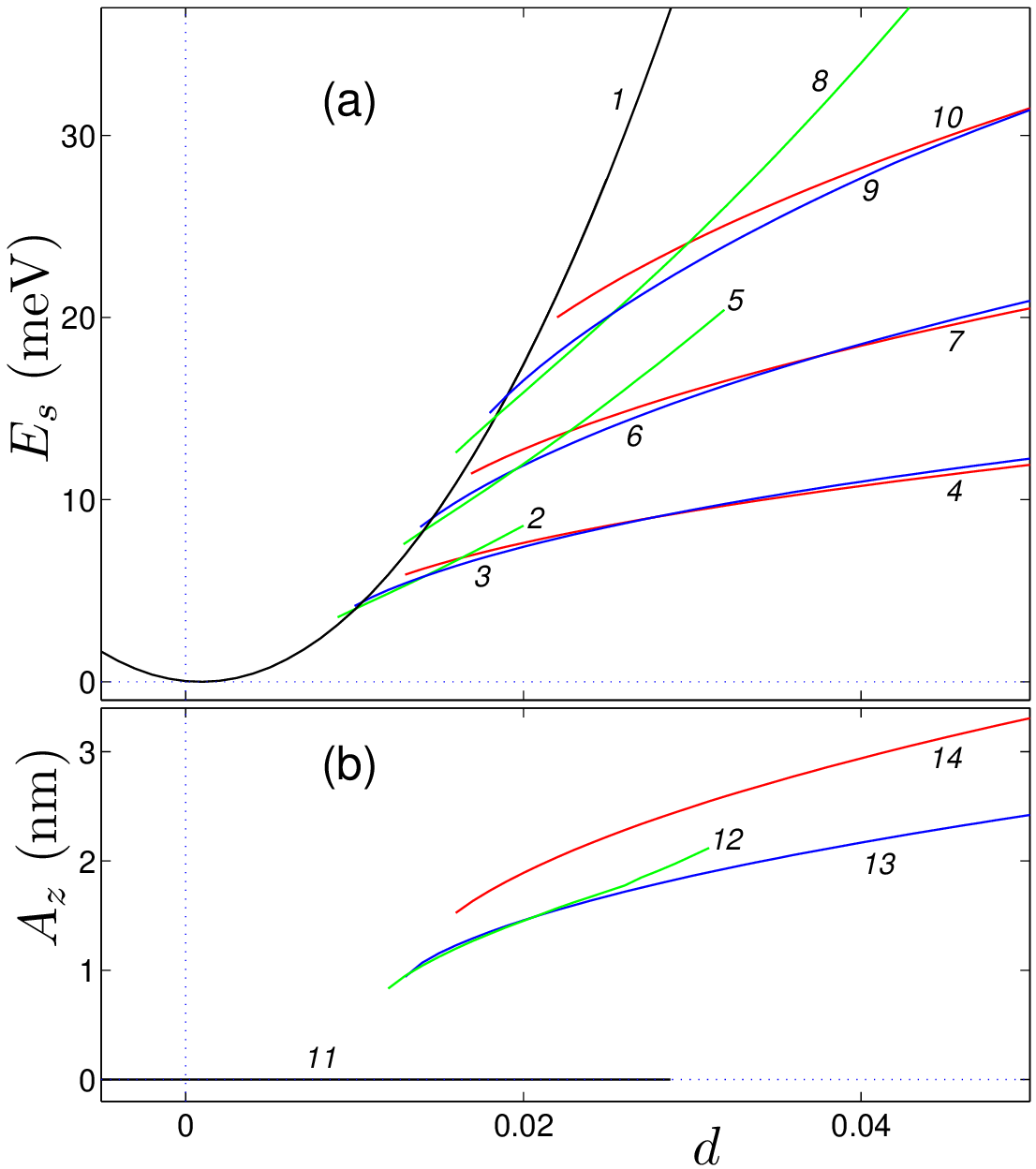}
\end{center}
\caption{\label{fg06}\protect
Dependence of (a) the specific energy $E_s=(E-E_0)/N$ and (b) the amplitude $A_z$ of the vertical displacements of atoms for the stationary state of a biaxially compressed graphene sheet on the dimensionless
coefficient of compression $d$: for a flat homogeneous state (curves 1, 11); for configuration with one vertical wrinkle (curves 2, 5, 8, 12); for a system of linear orthogonal non-intersecting (curves 3, 6, 9, 13) and  intersecting wrinkles (curves 4, 7, 10, 14).
Curves 2, 3, 4 were obtained at the energy of interaction with the substrate $\epsilon_0=0.032$~eV;
curves 5, 6, 7, 12, 13, 14 -- at $\epsilon_0=0.075$~eV;
curves 8, 9, 10 -- at $\epsilon_0=0.15$~eV.
}
\end{figure}
\begin{figure}[tb]
\begin{center}
\includegraphics[angle=0, width=1.0\linewidth]{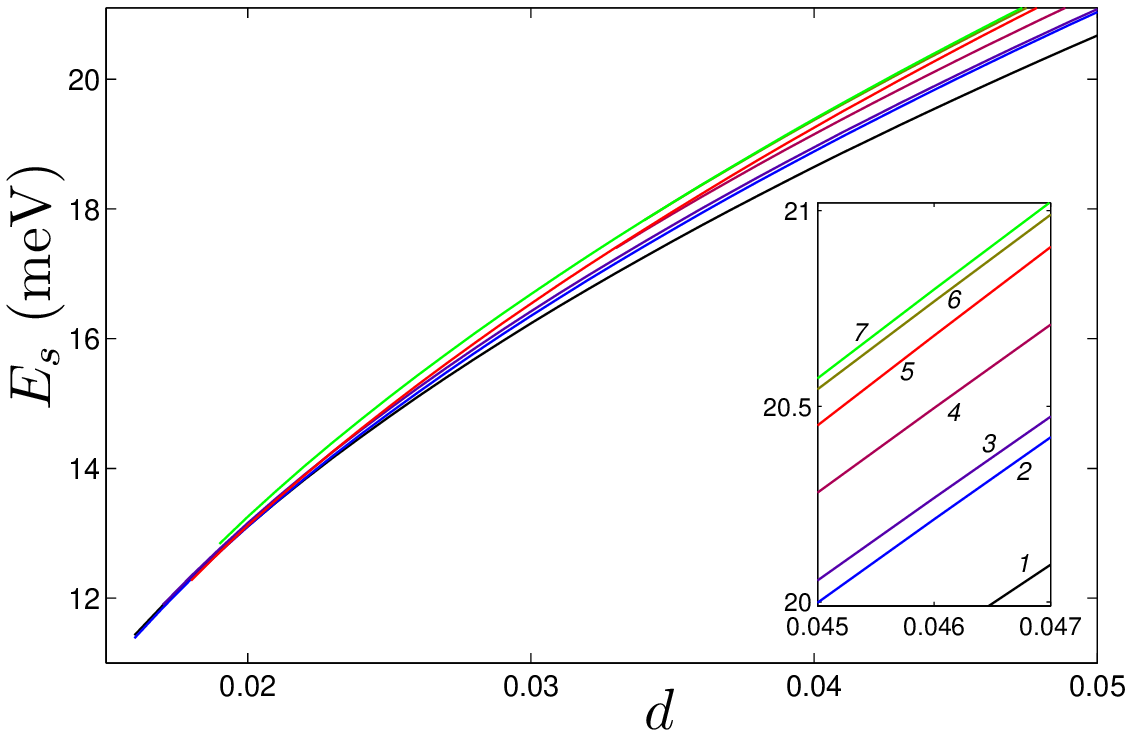}
\end{center}
\caption{\label{fg07}\protect
The dependence of the specific energy $E_s$ on the dimensionless biaxial compression coefficient $d$
for stationary states with orthogonal wrinkles, which are presented in Fig. \ref{fg04} (c) and \ref{fg05}.
Curve 1 corresponds to two orthogonal wrinkles with simple intersection,
the domain of existence $d>0.016$;
curve 2 -- to wrinkle network on Fig. \ref{fg05}(a), $d>0.011$;
curve 3 -- to the structure on panel (b), $d>0.016$;
4 -- (c), $d>0.032$;
5 -- (d), $d>0.017$;
6 -- (e), $d>0.033$;
7 -- (f), $d>0.018$.
In all cases adhesion energy $\epsilon_0=0.075$~eV.
}
\end{figure}

\section{Stationary states of a biaxially compressed sheet  \label{sec5}}

With biaxial compression ($d_x=d_y=d>0$), the number of possible stationary states of the compressed sheet
increases significantly. In addition to the uniformly compressed flat state of the sheet (stable at $d<d_2$),
there are also states with one linear wrinkle, with two linear orthogonal non-intersecting
or intersecting wrinkles -- see Fig. \ref{fg04}. Structures of the nanosheet with two orthogonal
wrinkles differ in topology of intersection area (with different structures of narrow flat valleys
in the intersection area) -- see Fig. \ref{fg05}.

The dependence of the specific energy $E_s$ and the amplitude of vertical displacements $A_z$ on the dimensionless coefficient of biaxial compression $d$ for the first four types of stationary states is shown in Fig.~\ref{fg06}, and for the rest -- in Fig. \ref{fg07}.
The energy of a flat uniformly compressed state is proprotional to square of compression
($E_s\propto d^2$ for $d\nearrow$), the energy of the single-wrinkle state increases in proportion to compression
($E_s\propto d$), and for states with two wrinkles, it is proportional to the square root of
the compression ($E_s\propto d^{1/2}$). The energy grows the slowest for a state with two intersecting
wrinkles -- see Fig. \ref{fg06}.
Therefore, there are three characteristic values for biaxial compression
$0<\beta_1<\beta_2<\beta_3$.
The type of the most energetically favorable graphene structure depends on compression coefficient $d$.
Under weak compression $d<\beta_1<d_2$ it is a flat uniformly compressed state; at $\beta_1<d<\beta_2$ it is single-wrinkled state;  at $\beta_2<d<\beta_3$ -- two orthogonal non-intersecting wrinkles and with strong compression $d>\beta_3$ -- a state with two intersecting orthogonal wrinkles.
The amplitude of vertical displacements $A_z$ for a single-wrinkle nanosheet is proportional to compression,
and for states with two orthogonal wrinkles - in proportion to the square root of compression ($A_z\propto
d^{1/2}$ for $d\nearrow$) -- see Fig.~\ref{fg06} (b).
The dependence of these characteristic values of biaxial compression on the adhesion energy
$\varepsilon_0$ is presented in the Table \ref{tab4}. The larger the $\varepsilon_0$, the larger
these values become.

States with two orthogonal wrinkles with more complex topology in the area of their intersection
(see Fig. \ref{fg05}) always have higher energy than a state with simple
wrinkle crossing structure -- see Fig. \ref{fg07}.

The states of a biaxially compressed graphene sheet with orthogonal non-intersecting wrinkles (Fig.~\ref{fg04}~c)
were first modeled in the work \cite{Zhang2014} (low energy T-junctions), and the states with intersecting
orthogonal wrinkles (Fig.~\ref{fg04}~b and \ref{fg05}~c) -- in Ref. \cite{Li2016}.
\begin{figure}[tb]
\begin{center}
\includegraphics[angle=0, width=1.0\linewidth]{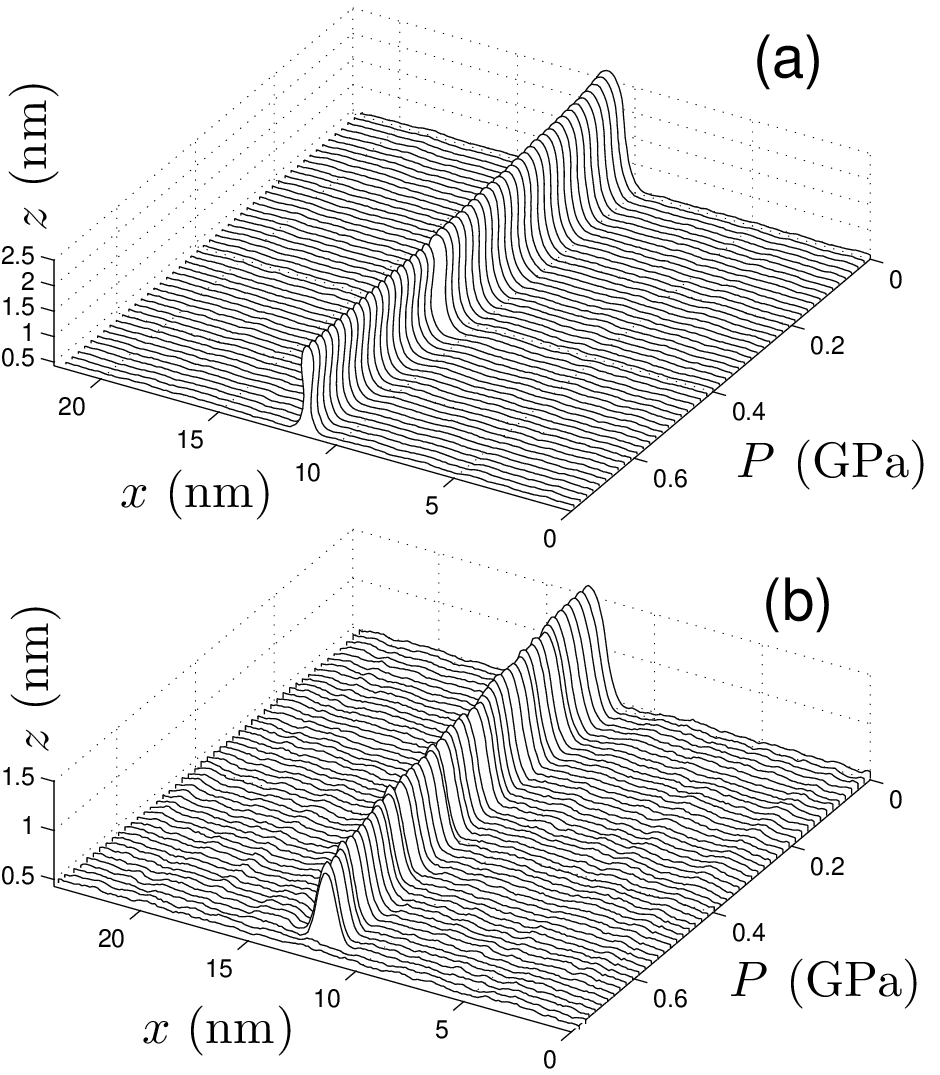}
\end{center}
\caption{\label{fg08}\protect
The change in the cross-section of a continuous wrinkle with increasing pressure $P$ at (a) uniaxial
${\bf d}=(0.06,0)$ and (b) biaxial compression ${\bf d}=(0.017,0.017)$ of a graphene sheet
(the energy of interaction with the substrate $\epsilon_0=0.075$~eV, temperature $T=300$~K).
With uniaxial compression at a pressure of $P_c=0.44$~GPa, the wrinkle collapses into a vertical fold.
With biaxial compression at a pressure of $P_f=0.72$~GPa, the wrinkle is flattened (the sheet turns into a flat state).
}
\end{figure}
\begin{figure}[tb]
\begin{center}
\includegraphics[angle=0, width=1.0\linewidth]{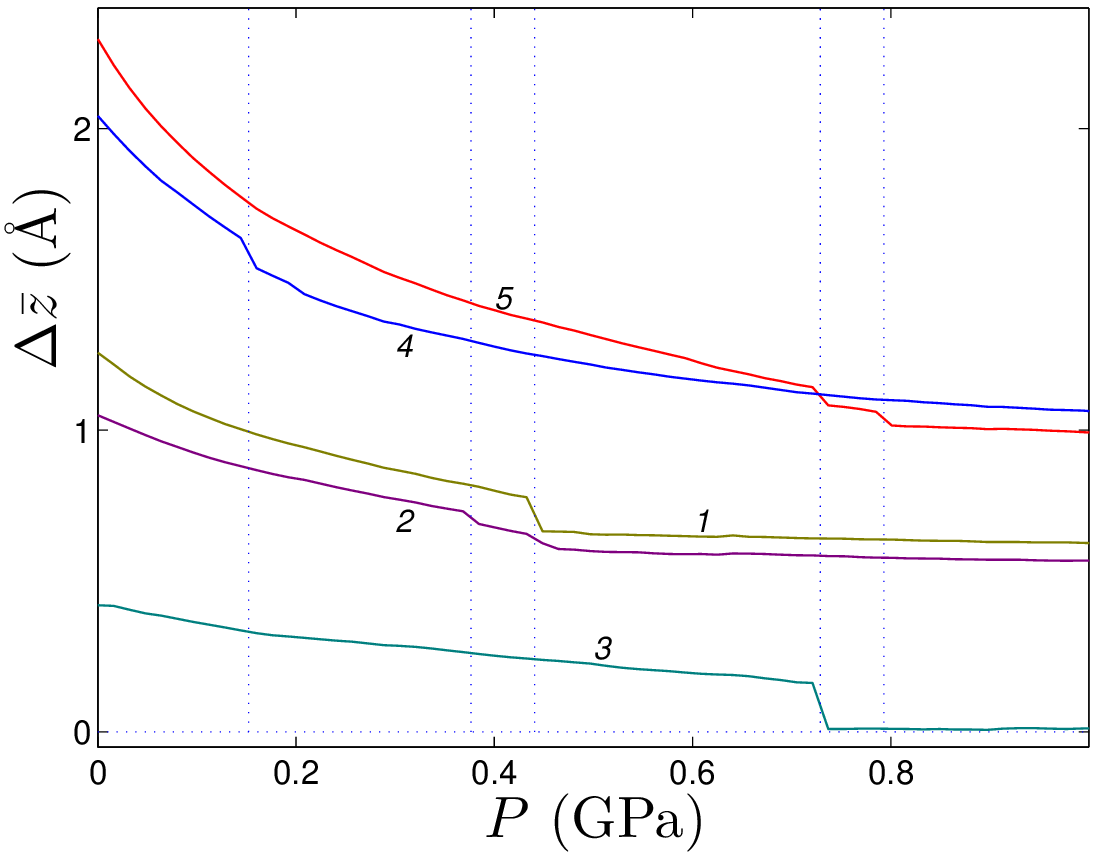}
\end{center}
\caption{\label{fg09}\protect
Dependence of the magnitude of the additive vertical displacements of the sheet atoms $\Delta\bar{z}$ on the pressure $P$ for the following initial states: a sheet compressed along one axis with a continuous wrinkle [compression ${\bf d}=(0.06,0)$, curve 1] and with connected wrinkles of finite length $L=A_y^0/2$ (curve 2); a biaxially compressed sheet with one continuous wrinkle [${\bf d}=(0.017,0.017)$, curve 3], with two orthogonal non-intersecting [${\bf d}=(0.05,0.05)$, curve 4] and intersecting wrinkles (curve 5).
The vertical dotted lines correspond to pressure values $P=0.152$, 0.376, 0.440, 0.720, and 0.793~GPa.
}
\end{figure}

\section{Stability of stationary states to thermal fluctuations  \label{sec6}}

Note that all stationary states of the sheet obtained by solving the minimum
energy problem are always stable solutions.
To test the stability of these states to thermal fluctuations, we will simulate the dynamics
of a sheet at temperatures $T=600$ and 900~K. To do this, we numerically integrate
the system of equations of motion (\ref{f7}) with an initial conditions corresponding to the stationary state
$\{{\bf u}_n(0)={\bf v}_n,~\dot{\bf u}_n(0)=0\}_{n=1}^N$.
For certainty, let us take the medium value of the energy of interaction with the substrate $\varepsilon_0=0.075$~eV.

Numerical simulation of the dynamics showed that at $T=600$~K, the stationary states of uniaxially compressed sheet (Sec. \ref{sec4}, Fig. \ref{fg02}) are resistant to thermal fluctuations. For example, configuration with one continuous wrinkle is stable at $d>0.022$ and the nanosheet with several adjacent wrinkles of finite length is stable at $d>0.024$.
At small values of compression, thermal fluctuations can cause wrinkles of finite length to join, while at larger values, they retain initial shape.
For instance, at $T=900$~K, a stationary state with wrinkles of length $L=A_y^0/2$ transits to a flat uniformly compressed state at $d<0.028$, while it goes to a state with one continuous wrinkle at $0.028\le d<0.033$ and retains its initial shape at $d\ge 0.033$.

With biaxial compression $d_x=d_y=d>0$, all the described stationary states of the sheet (Sec. \ref{sec5})
are resistant to thermal fluctuations at sufficiently large values of $d$, and at small values
they can transition to more energetically favorable states.
So, at $T=600$~K, the nanosheet with two intersecting orthogonal
wrinkles is flattened at $d<0.017$, transitions into configuration with one vertical wrinkle at $0.017\le d \le 0.019$, while retaining its shape at $d\ge 0.020$.
The nanosheet with two non-intersecting orthogonal wrinkles is flattened at $d<0.017$, while retaining its shape at $d>0.017$.
Note that there is no transition between states with intersecting and non-intersecting orthogonal wrinkles due to the difference in their topology. The other structures shown in Fig. \ref{fg05} also remain stable under strong compression.

\section{The effect of pressure on the shape of wrinkles \label{sec7}}

Dynamics modeling also showed that all stationary structures with wrinkles
obtained at zero pressure are also preserved at $P>0$. An increase in pressure
leads to a monotonous decrease in the size of wrinkles (to a decrease in the free space volume
under them) -- see Fig.~\ref{fg08}.
When the critical pressure value $P_f$ is reached, the wrinkle
either completely flattens (disappears), turning the sheet into a flat state, or at pressure $P_c$
it collapses, turning into the form of a vertical two-layer fold of the sheet.
So, with uniaxial compression $d=0.06$ and temperature $T=300$~K, the linear wrinkle
collapses at pressure $P_c=0.44$~GPa, and with biaxial compression $d=0.017$, the wrinkle disappears at $P_f=0.72$~GPa.
\begin{figure}[tb]
\begin{center}
\includegraphics[angle=0, width=1.0\linewidth]{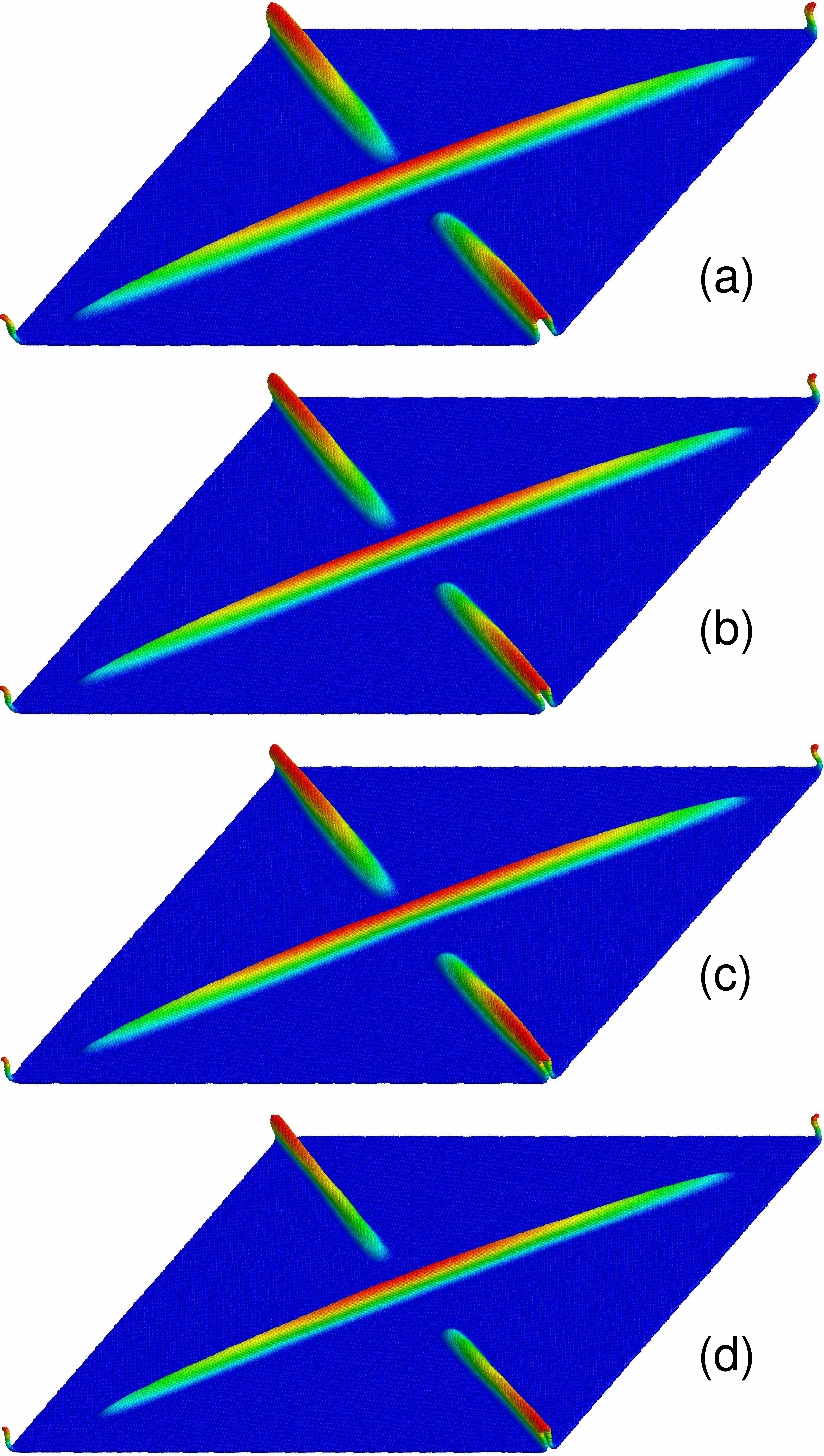}
\end{center}
\caption{\label{fg10}\protect
Structure of biaxially compressed nanosheet with two orthogonal non-intersecting wrinkles under pressure
(a) $P=0$, (b) 0.144, (c) 0.160 and (d) 0.736 GPa
(compression $d=0.05$, $\epsilon_0=0.075$~eV, $T=300$~K).
One unit cell is shown.
}
\end{figure}
\begin{figure}[tb]
\begin{center}
\includegraphics[angle=0, width=0.99\linewidth]{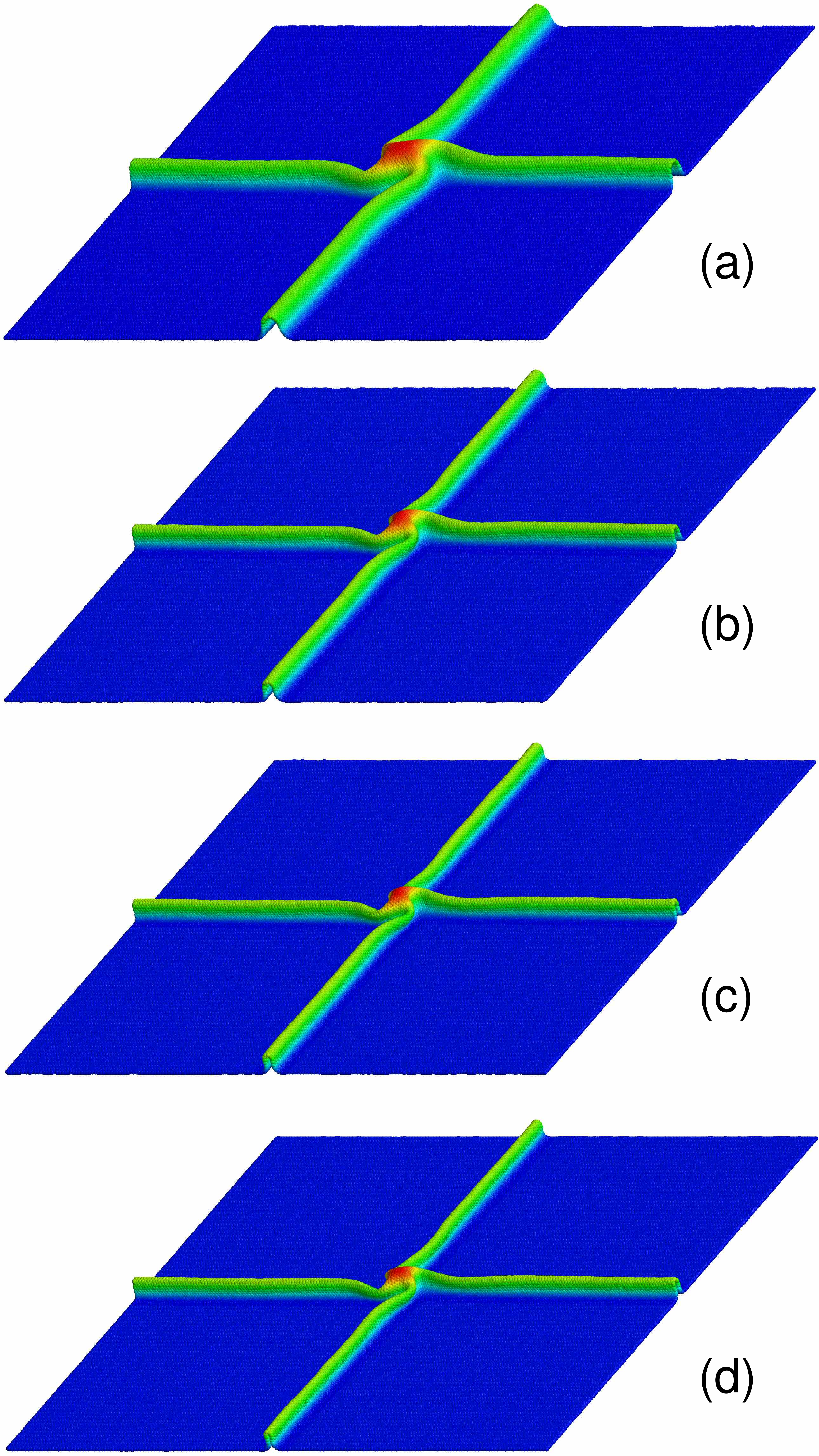}
\end{center}
\caption{\label{fg11}\protect
Structure of biaxially compressed nanosheet with two orthogonal intersecting wrinkles 
under pressure (a) $P=0$, (b) 0.368, (c) 0.720 and (d) 0.736 GPa
(compression $d=0.05$, $\epsilon_0=0.075$~eV, $T=300$~K).
One unit cell is shown.
}
\end{figure}

It is convenient to monitor the change in the shape and structure of wrinkles using an average value of
$z$ coordinates of the atoms of the sheet
$$
\bar{z}=\frac{1}{N}\sum_{n=1}^N z_n
$$
(this value characterizes the amount of space between the graphene sheet and the substrate surface).

The value of $\bar{z}$ always decreases monotonously with increasing pressure.
Even for an uncompressed flat sheet $\bar{z}\searrow$ at $P\nearrow$ (under the influence of increasing
pressure, the sheet presses more and more strongly against the flat substrate $z\le 0$).
It is convenient to determine the amount of additional vertical displacements of sheet atoms associated with
the presence of wrinkles as $\Delta\bar{z}=\bar{z}-\bar{z}_0$, where $\bar{z}_0$ is the average value
$z$ coordinates for an uncompressed (flat) sheet.
The dependence of $\Delta\bar{z}$ on the pressure of $P$ is shown in Fig. \ref{fg09}.
The value of $\Delta\bar{z}$ decreases monotonously with increasing pressure, or remains constant.
A sharp decrease in $\Delta\bar{z}$ indicates structural changes of wrinkles.
Thus, with uniaxial compression ${\bf d}=(0.06,0)$ at $P_c=0.44$~GPa, the
wrinkle collapses into a vertical two-layer fold (curve 1). After the collapse of the wrinkle
(at pressure $P>P_c$) the value of additional vertical displacements $\Delta\bar{z}$
practically does not change. For a system of connected wrinkles of finite length, the collapse
of wrinkles begins at their centers and then spreads to their ends (collapse
occurs at $0.376<P\le 0.44$~GPa, curve 2).

With biaxial compression ${\bf d}=(0.017,0.017)$, the wrinkle is completely flattened at $P_f=0.72$~GPa and
at $P>P_f$ there are no additional vertical displacements ($\Delta\bar{z}=0$, curve 3).

Let us consider nanosheet with two orthogonal non-intersecting wrinkles under compression ${\bf d}=(0.05,0.05)$. When the external pressure increases, a sharp change in $\Delta\bar{z}$ occurs at
$P_c=0.152$~GPa (curve 4). It corresponds to a partial collapse in the central part of wrinkles into a vertical fold (see Fig.~\ref{fg10}). Therefore, at $P>P_c$, a monotonous
decrease of $\Delta\bar{z}$ continues. As for the nanosheet with two orthogonal
intersecting wrinkles, a sharp change in $\bar{z}$ occurs at two pressure values:
$P_{c,1}=0.729$ and $P_{c,2}=0.793$~GPa (curve 5). At the first value, the non-overlapping areas
of wrinkles collapse, and at the second one, the area of the intersection of wrinkles collapses
-- see Fig.~\ref{fg11}. A further increase in pressure does not lead to a change in the value
$\Delta\bar{z}$ (all wrinkles are fully compressed and therefore do not change their shape).

\section{Conclusion  \label{sec8}}

Using an all-atom model, numerical modeling of the formation of wrinkles in a
graphene sheet lying on a flat substrate during its compression along one and two axes has been carried out.
It is shown that under uniaxial compression the nanosheet can transition into several stable wrinkled states: the most energetically favorable one is a linear wrinkle of infinite length. Higher energy states include wrinkles of finite length aligned along the same line where their ends partially overlap (wrinkles ends are separated by a narrow flat valley-like sections of the nanosheet).

With equal biaxial compression, the number of possible stationary states of the wrinkled sheet increases significantly.
The graphene nanosheet can contain one linear wrinkle or two linear orthogonal
to each other non-intersecting and intersecting wrinkles. Configurations with intersecting wrinkles may
have different structures of intersection area. For biaxial compression, there are three characteristic
compression values $0<\beta_1<\beta_2<\beta_3<1$. These values delineate different the most energetically favorable structures of the nanosheet. A flat uniformly compressed state is favorable at $d<\beta_1$, a state with one linear wrinkle -- at $\beta_1<d<\beta_2$ and the nanosheet with two orthogonal non-intersecting wrinkles -- at $\beta_2<d<\beta_3$. Under strong compression $d>\beta_3$ the nanosheet with two intersecting wrinkles is the most favorable.

All stationary wrinkled states of the nanosheet are resistant to thermal fluctuations at temperatures of $T=600$~K and above.
Only at small values of uniaxial compression $d<0.033$ the bound states of wrinkles of finite length can unite into one linear wrinkle under the action of thermal oscillations.

External pressure does not change the wrinkle structure of the compressed sheet.
An increase in pressure leads to a monotonous decrease in the size of wrinkles.
It is shown that there is always a critical pressure value at which the wrinkle either completely
flattens (disappears), turning the sheet into a flat state, or collapses, taking
a denser form of a vertical two-layer fold. The first scenario is possible only
with weak compression, when the wrinkles of the nanosheet have a small height.

\section*{Acknowledgements}

Computational facilities were provided by the
Joint Supercomputer center (JSCC) of the National Research Center "Kurchatov Institute".
This research was supported by the Program of Fundamental Research of the Russian Academy of Sciences (project FFZE-2022-0009).

\end{document}